\documentclass[11pt]{article}
\usepackage{amsmath,amsthm,amssymb,amscd}
\usepackage{hyperref}
\usepackage{geometry}
\usepackage{tikz-cd}
\usepackage{booktabs} 
\usepackage{graphicx} 
\usepackage{multirow}
\usepackage{tikz}
\usetikzlibrary{calc,positioning,3d}
\usepackage[normalem]{ulem} 
\usepackage{imakeidx}
\usepackage{makecell}
\usepackage{float}
\makeindex

\usepackage{xr-hyper}
\usepackage{cleveref}   

\usepackage{subfigure} 

\theoremstyle{definition}

\title{CAP: Commutative Algebra Prediction of Protein-Nucleic Acid Binding Affinities}
		
	\author{Mushal Zia$^1$,  
	Faisal Suwayyid $^{1,2}$, 
	Yuta Hozumi\footnote{Current address: School of Mathematics, Georgia Institute of Technology, Atlanta, GA, USA.} ~$^1$\\
	JunJie Wee$^1$, 
	Hongsong Feng$^3$,
		and Guo-Wei Wei\footnote{Corresponding author: Guo-Wei Wei (weig@msu.edu).~}$^{~1,4,5}$ \\		
		$^1$Department of Mathematics,\\
		Michigan State University, MI 48824, USA.\\
		$^2$Department of Mathematics,\\
		King Fahd University of Petroleum and Minerals, Dhahran 31261, KSA.\\
		$^3$Department of Mathematics and Statistics,\\
		University of North Carolina at Charlotte, Charlotte, NC 28223, USA\\
		$^4$Department of Electrical and Computer Engineering,\\
		Michigan State University, MI 48824, USA.\\
		$^5$Department of Biochemistry and Molecular Biology,\\
		Michigan State University, MI 48824, USA.
	}

\date{} 
\begin{document}
	\maketitle 
	
\begin{abstract}	    

An accurate prediction of protein-nucleic acid binding affinity is vital for deciphering genomic processes, yet existing approaches often struggle in reconciling high accuracy with interpretability and computational efficiency. In this study, we introduce commutative algebra prediction (CAP), which couples persistent Stanley-Reisner theory with advanced sequence embedding  for predicting protein-nucleic acid binding affinities. CAP encodes proteins through transformer-learned embeddings that retain long-range evolutionary context and represents DNA and RNA with $\textit{k}$-mer algebra embeddings derived from persistent facet ideals, which capture fine-scale nucleotide geometry. We demonstrate that CAP surpasses the SVSBI protein-nucleic acid benchmark and, in a further test, maintains reasonable performance on newly curated protein-RNA and protein-nucleic acid datasets. Leveraging only primary sequences, CAP generalizes to any protein-nucleic acid pair with minimal preprocessing, enabling genome-scale analyses without 3D structural data and promising faster virtual screening for drug discovery and protein engineering.

 \end{abstract}
	
  
 \newpage	
{\setcounter{tocdepth}{4} \tableofcontents}
	 \setcounter{page}{1}
	 \newpage	
	
\section{Introduction}

Protein-nucleic acid recognition underpins a spectrum of essential biological processes. These processes form the backbone of cellular processes central to life, ranging from DNA replication, transcription, genome stability to RNA transport, splicing, and post-transcriptional regulation. An accurate modeling of protein-DNA and protein-RNA binding is crucial not only for fundamental biology but also for therapeutic innovations, accelerating genomic medicine, synthetic biology, and drug design \cite{bertoldo2023rna}. Molecular-level mechanisms are mediated by proteins through recognization and binding to specific DNA or RNA sequences. Such mechanisms includes hydrogen bonding, electrostatic attraction, hydrophobic interaction, physicochemical force, and structural complementarity. Disruption in such binding events can lead to the development of various diseases such as, cancer, neurogenerative conditions, and autoimmune disorders \cite{kisby2024introducing}. Thus, the orchestration of these associations is crucial not only for decoding biomolecular functions but also for paving the way to new therapies and drug discovery. 

A variety of conventional techniques have been adopted in past for the gold-standard measurement of binding affinity predictions of protein-protein/ligand systems, like fluorescence spectroscopy, electrophoretic mobility shift assays (EMSA), surface plasmon resonance (SPR), isothermal titration calorimetry, and filter binding \cite{stockley2009filter}. However, they are typically resource-intensive and time consuming. To address these limitations, a spectrum of in silico   strategies have been developed. Classical physics-based methods such as thermodynamic integration, free energy perturbation, molecular Mechanics Poisson-Boltzmann Surface Area (MM-PBSA) offer detailed estimates but are particularly laborious for large biomolecular systems. Non-classical approaches include but not limited to knowledge-driven potentials \cite{zhang2005knowledge}, empirical scoring functions \cite{nithin2019structure}, force-field scoring functions \cite{yin2008medusascore}, and machine-learning approaches using engineered descriptors \cite{bitencourt2019machine}. These methods have delivered promising results in protein-protein or protein-ligand systems. However, they tend to underperform on DNA owing to its unique conformational complexities and lack of availability of comprehensive binding affinity datasets. 

On the other hand, various experimental and computational approaches have been applied over the past decade to understand protein-RNA interactions, ranging from high-throughput CLIP-seq \cite{hafner2010transcriptome}, purely seq-based prediction algorithms  \cite{zhao2011structure,liu2016prediction,shen2023svsbi} to knowledge-based scoring \cite{tuszynska2011dars} and coarse-grained docking \cite{setny2011coarse}. In addition, resolving protein-RNA complex structures through X-ray crystallography or nuclear magnetic resonance (NMR) is hampered by inherent flexibility of many RNA partners which makes such determinations technically challenging and prohibitively slow. Although several protein-RNA docking approaches have substantially accelerated the discovery of RNA-protein interaction sites and expanded available structural decoys, yet none provide quantitative binding affinity measurements, leaving a paucity of RNA affinity data \cite{iwakiri2016improved}.

In light of these limitations, data-driven models, particularly those built on machine-learning (ML) frameworks, have recently reshaped computational molecular biology by emerging as an integral components for modern drug design \cite{lin2023evolutionary,song2024multiobjective}. The integration of bioinformatics \cite{lo2018machine} with ML and deep learning, when applied to protein-nucleic acid systems, has demonstrated promising shift by enabling the prediction of molecular interactions in a complex high-dimensional data at an unprecedented scale. Recently, mathematical artificial intelligence (AI), particularly topological deep learning (TDL) first introduced in 2017 \cite{cang2017topologynet}, has emerged as a new paradigm for rational learning in data science \cite{papamarkou2024position}, leading to a new frontier for modeling intricate biomolecular systems\cite{nguyen2020review}, such as proteins and their interactions \cite{rana2023geometry,wang2025join}. Notably, mathematical AI secured victories in D3R Grand Challenges, an annual worldwide competition series in computer-aided drug design \cite{nguyen2019mathematical,nguyen2020mathdl}. Recently, topological sequence analysis (TSA) has been proposed as a viable approach for biological sequence modeling \cite{hozumi2024revealing}. Delta complex approaches of TSA enable the efficient treatment of large genomic sequences \cite{liu2025topological2}, while category theory approaches of TSA achieve higher accuracy for nearly identical genetic variants \cite{liu2025topological}. 

Some early studies laid important groundwork for protein-DNA binding interactions using ML frameworks, such as Zhao et al.’s atomic pairwise statistical potential (DDNA3) \cite{zhao2010structure}, Rastogi et al.’s sequencing-based affinity profiling \cite{rastogi2018accurate}, and Barissi et al.’s physics-based machine-learning method (DNAffinity) \cite{barissi2022dnaffinity}. Building on these advances, subsequent models such as PreDBA \cite{yang2020predba}, PDA-Pred \cite{harini2023pda}, and emPDBA \cite{yang2023empdba} have broadened the affinity-prediction toolbox. Likewise, for protein-RNA interactions, Yang et al. \cite{yang2013dataset} assembled the first quantitative dataset of protein-RNA affinities, which was later followed by the development of several structure-driven learning frameworks, each developed with its own curated dataset, including the methods proposed by Nithin et al. \cite{nithin2019structure} and PredPRBA \cite{deng2019predprba}, as well as models including Hong et al. \cite{hong2023updated} and the more recent PRA-Pred \cite{harini2024pred}. Though these advancements have pushed the field forward, yet most approaches rely on stacked regressors, class-specific framework architectures to curb overfitting in small datasets, and basic interface metrics. 

Commutative algebra is a mathematical study of commutative rings, their ideals, modules, and related structures \cite{eisenbud2013commutative}. As a foundational framework in pure mathematical fields such as algebraic number theory, homological algebra, and algebriac geometry, commutative algebra has seen limited application in data science and the biological sciences. In a recent effort, Suwayyid and Wei introduced persistent Stanley-Reisner theory (PSRT) to forge a connection between commutative algebra, algebraic topology, data science, and machine learning \cite{suwayyid2025persistent}. This method has already yielded promising results in protein-ligand binding affinity predictions via a structure-based approach \cite{feng2025caml}. 

The Stanley-Reisner theory investigates the commutative algebra of simplicial complexes, assemblies of vertices, edges, triangles, and their higher-order faces, through the study of square-free monomial ideals in polynomial rings \cite{stanley2007combinatorics,francisco2014survey}.  PSRT  integrates these notions with multiscale analysis, enabling commutative algebra learning and predictions. PSRT examines the evolution of Stanley-Reisner ideal across a filtration, yielding a range of algebraic and topological invariants, such as persistent graded Betti numbers (computed via Hochster’s formula),   persistent f- and h-vectors, and persistent facet ideals. In particular, facet-persistence barcodes which document the “birth” and “death” of each facet ideal during filtration have been developed for real-world data applications. Thus, PSRT 
is a new multiscale approach  
that offers rich geometric, topological, and combinatorial analysis not readily accessible to conventional mathematical, statistical, and physical approaches.

In this study, we investigate commutative algebraic predictions (CAPs) for the sequence-based modeling of protein-nucleic acid binding affinities, a direction not previously explored. Our CAP offers algebraic sequence analysis (ASA), an algebra-based sequence-driven alternative to conventional structure-based approaches, facilitating an effective modeling of biomolecular interactions using primary sequence information alone. We employ a dataset assembled by Shen \textit{et al.}~\cite{shen2023svsbi}, comprising 186 protein-nucleic acid complexes derived from the PDBbind v2020 database, referred to here as S186. Each complex in this dataset is represented by pairing rich residue-level embeddings of the protein sequence with nucleotide-level descriptors of the associated nucleic acid strand (DNA or RNA). These complementary sequence-based features implicitly capture interactions such as hydrogen bonding, electrostatic contacts, base stacking, and van der Waals forces. Additionally, we evaluate CAP on two newly curated datasets: a protein-RNA dataset (S142) containing 142 complexes, and a protein-nucleic acid dataset (S322) containing 322 complexes. The results demonstrate CAP’s robust predictive performance, highlighting a promising new approach for affinity prediction without explicit reliance on structural models. 

In the following sections, we present our results followed by a brief discussion on existing methodologies, the two new datasets, and feature extraction processes with a detailed demonstration of robust predictive performance of CAP for protein-nucleic acid binding affinity. We also elaborate on the mathematical foundations and computational interpretability of CAP. Finally, we conclude by discussing our findings and outlining potential future research directions.

\section{Results and Discussion} \label{sec:Results}

\subsection{Overview of the CAP model} 

\begin{figure}[t!]
	\centering
	\subfigure{
		\includegraphics[width=\textwidth]{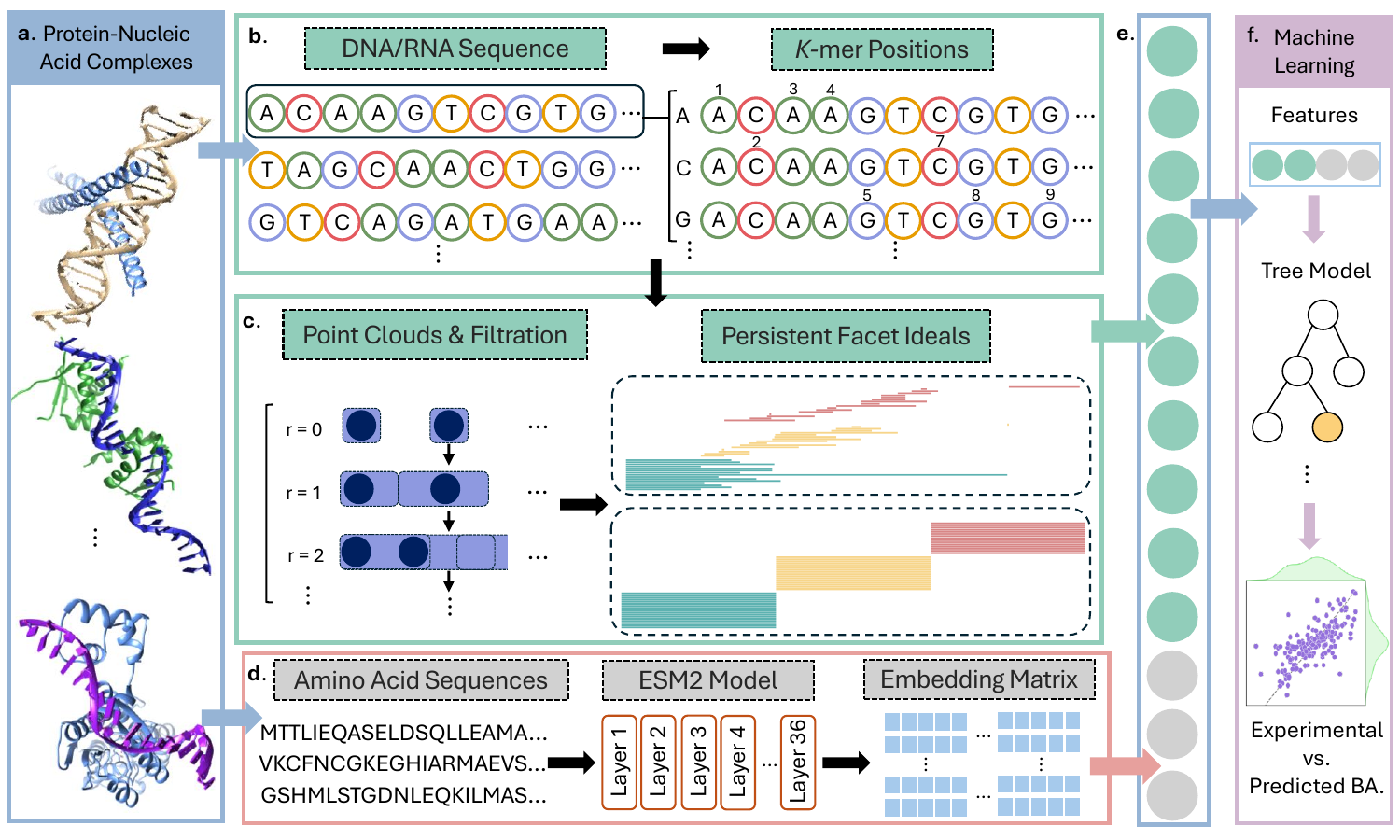}
	}
	\caption{Illustration of the workflow in CAP. (a) Protein-nucleic acid complexes. (b) $\textit{k}$-mers are extracted from the DNA sequences of the given complexes. For each $\textit{k}$-mer=1, the set of its occurrence positions within the sequence is treated as an input data. (c) The persistent facet ideals associated with these input data for three dimensions are then computed and used as topological features for the corresponding $\textit{k}$-mers. (d) From the  amino acid sequence for the given complexes, 2560 embedding vectors are generated using state-of-the-art ESM2 model with 36 layers. (e) The feature vectors of all DNA and proteins are concatenated to construct a genome-level topological representation. (f) Finally, our CAP model is trained using these features with machine learning techniques for the prediction task.} 
	\label{fig:concept}
\end{figure}

Figure \ref{fig:concept} illustrates the workflow of the proposed CAP model. For a given protein-nucleic acid complex (a), the nucleic acid sequence is processed with PSRT (b-c)to provide a CAP feature vector. Meanwhile, the protein sequence is processed with the ESM2 model to result in an embedding matrix (d). The combined feature vector (e) is fed into a machine learning model (f). More details of protein sequence embedding, nucleic acid sequence analysis, and machine learning parametrization can be found in the methods section.

\subsection{Protein-Nucleic Acid Binding Affinity Prediction}

\begin{figure}[t!]
	\centering
	\subfigure{
		\includegraphics[width=\textwidth]{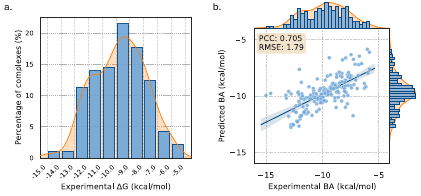}
	}
	\caption{(a) Distribution of experimental binding affinity $(\Delta G)$ in the S186 dataset. (b) A comparison between the experimental binding affinities and the predicted values from our CAP model for this dataset. The corresponding Pearson correlation coefficient (PCC) value is 0.705 with an RMSE of 1.79 kcal/mol.}
	\label{fig:scatterplot}
\end{figure}
 
A vital category of biomolecular interactions involves protein-nucleic acid binding which underpins essential cellular processes such as, catalysis, molecular transport, signal transduction, transcription, and translation. Moreover, these interactions preserves chromosomal integrity and regulates gene expression however, can contribute to pathologies including cancer, genetic disorders, autoimmune dieseases. Understanding how hydrogen bonds, van der Waals contacts, dipole-dipole interactions, electrostatics, van der Waals attractions, and hydrophobicity contribute to binding affinities then informs the rational design of new therapeutics such as structure-based drug design. In this study, we apply our CAP framework to capture these interactions and predict protein-nucleic acid binding affinity. 

We begin by evaluating the predictive accuracy of our PSRT-guided framework on the S186 dataset against the SVSBI model on protein-nucleic acid binding affinity \cite{shen2023svsbi}. Due to a lack of established benchmarks, Shen \textit{et al.} \cite{shen2023svsbi} assembled a dataset of 186 protein-nucleic acid complexes. In order to construct a high-quality dataset from PDBbind-v2020, the authors applied four stringent filters: 
(i) a complex was retained only if it contained one unique protein sequence and one unique nucleic acid sequence. Furthermore, multiple chains carrying the same sequence were allowed, but any chain with mixed or hybrid bases (e.g., both T and U) was discarded; 
(ii) only complexes with experimentally determined binding affinities measured at 298 K were kept; 
(iii) entries containing ambiguous labels (\texttt{\textasciitilde}, \texttt{<}, or \texttt{>}) were not considered; 
(iv) exclude those complexes that whose nucleic acid sequence length is fewer than 5. 

The SVSBI framework leverages an ESM-based Transformer for proteins and DNABERT for nucleic acids and records an average Pearson of 0.669 and RMSE of 1.98 kcal/mol. On the other hand, CAP achieves robust prediction performance with a Pearson correlation coefficient \(PCC\) of 0.705 with RMSE of 1.79 kcal/mol, surpassing the existing SVSBI model \cite{shen2023svsbi} for the protein-nucleic binding affinity prediction on the same dataset as shown in Table~\ref{table:PCC1}. Furthermore, we analyze the distribution of experimental binding affinity $\Delta G$ values of S186 as shown in Figure~\hyperref[fig:scatterplot]{\ref*{fig:scatterplot}a}. We also observe that $\Delta G$ ranges from -15 kcal/mol to -6 kcal/mol and 66\% of the complexes have $\Delta G$ of -11 to -7 kcal/mol. An illustrative comparison between the experimental binding affinities and the predicted values from our CAP model for this dataset is presented in Figure~\hyperref[fig:scatterplot]{\ref*{fig:scatterplot}b} and a full comparison of experimental and predicted binding affinities is provided in Table~\ref{tab:S186}. In addition, a representation of how binding free energy varies with sequence length in S186 is represented in Figure~\hyperref[fig:lenDist]{\ref*{fig:lenDist}a}. While the strand lengths range from 5 to 45$\sim$nt, we observe that the shortest strands (5$\sim$nt) bind weakly on average (mean $\Delta G=-7.77$ kcal/mol), whereas the longest strands (45$\sim$nt) reach an average $\Delta G=-9.57$ kcal/mol. Moreover, we also note that the length interval with the highest coverage (10-15$\sim$nt, 67 complexes) has a mean $\Delta G$ of $-9.32$ kcal/mol. Our CAP framework provides a powerful and structure-free method, even though that the size of our dataset is modest.

\begin{table}[htb!]
	\small
	\centering
	\caption{
		Comparison of prediction performance between existing SVSBI model and our model, CAP, on S186 for protein-nucleic acid binding affinity prediction. Reported metrics include Pearson correlation coefficient ($PCC$) and RMSE values in kcal/mol. All results are averaged over twenty independent runs with different random seeds, and the average metric values are reported.}
	\label{table:PCC1}
	\vspace{10pt}   
	\begin{tabular}{l | c | c}
		\hline
		\textbf{Model} & \textbf{PCC} & \textbf{RMSE (kcal/mol)} \\
		\hline
		SVSBI \cite{shen2023svsbi}   & 0.669  &  1.98\footnotemark \\ 
		\textbf{CAP}   & \textbf{0.705} & \textbf{1.79} \\
		\hline
	\end{tabular}
\end{table}

\footnotetext{%
	The original RMSE of 1.45 reported by Shen et al. \cite{shen2023svsbi} was not converted into kcal/mol; here we apply the factor 1.3633 to obtain 1.98 kcal/mol.}

Several existing protein-DNA affinity models \cite{zhao2010structure,yang2020predba,harini2023pda,yang2023empdba} and protein-RNA models \cite{deng2019predprba,hong2023updated,harini2024pred} have broadened the binding prediction toolbox. However, they depend on large, handcrafted feature sets and category-specific tuning steps that complicate workflows and have limit generalizability. CAP sidesteps these limitations by coupling persistent Stanley-Reisner theory with a concise sequence-level interaction embedding and a single gradient-boosting regressor. By summarizing residue-nucleotide contacts directly from the paired protein and nucleic-acid sequences, CAP removes the need to count hydrogen bonds, tally base-pair frequencies, model stacking ensembles, or train separate predictors for different structural subclasses. Because CAP relies solely on primary sequences, it can be applied to any protein-DNA/RNA pair without 3D structures, subclass labels, or bespoke feature engineering, offering higher predictive power with far less manual effort and providing an extensible framework for large-scale studies of protein-nucleic recognition.

\subsection{Discussion}\label{sec:Discussion}

Within the spectrum of existing binding affinity approaches for protein-DNA, the most recent is emPDBA \cite{yang2023empdba}, which integrates 106 sequence- and structure-based features into a stacking ensemble applied to 340 protein-DNA complexes at 40 \% protein sequence similarity. After subclassifying DNA by structural form and applying feature selection method, it reaches a correlation value of 0.66 with mean absolute error (MAE) = 1.24 kcal/mol, compared with 0.12 (MAE = 1.64 kcal/mol) on an unclassified set. Another model, PDA-Pred \cite{harini2023pda}, built on 117 features for each of the 391 complexes at 25 \% sequence identity, employs similar classifications of DNA structures and with an additional protein functional categorization followed by a feature selection method (jack-knife) to achieve \(PCC\) = 0.78, dropping to just \(PCC\) = 0.21 without any categorization. Furthermore, on a common independent testing set of 36 complexes, emPDBA showed an \(PCC\) = 0.53 with MAE = 1.11 kcal/mol, whereas earlier ensemble regression methods like PreDBA \cite{yang2020predba} attained \(PCC\) = 0.30, and the statistical potential DDNA3 \cite{zhao2010structure} only \(PCC\) = 0.09 with MAE = 2.05 kcal/mol and MAE = 1.80 kcal/mol, respectively. 

While numerous models exist for protein-DNA binding affinity prediction, there are only a limited number of models available for predicting protein-RNA binding affinity \cite{deng2019predprba,hong2023updated,harini2024pred}. The most recent, PRA-Pred model \cite{harini2024pred}, consists of 217 protein-RNA complexes with proteins clustered at a 25\% sequence identity cutoff, adopts a classification strategy similar to that used in protein-DNA binding affinity prediction. It categorizes RNA into five structural subclasses and proteins into functional categories and achieves a Pearson correlation of 0.77 and a mean absolute error (MAE) of 1.02 kcal/mol using a jack-knife feature selection strategy. On the standard blind set of 44 complexes, the model obtains an overall performance of \( PCC = 0.60 \) with an MAE of 1.47 kcal/mol. Another model, PRdeltaGPred \cite{hong2023updated}, starts with 63 interface-based features and applies a custom feature reduction process to improve prediction accuracy. Despite this tailored feature selection strategy, PRdeltaGPred reports \( PCC = 0.41 \) with an MAE of 1.83 kcal/mol on the standard 44-complex blind set. The third model, PredPRBA \cite{deng2019predprba}, is trained on 103 protein-RNA complexes filtered at a 40\% sequence identity cutoff. It makes use of both sequence-level and structural features at the whole-complex level, and applies jack-knife test within six predefined RNA subclasses. While the model’s overall performance on the standard 44-complex blind test is limited (\( PCC = 0.07 \), MAE = 2.07 kcal/mol, its best pre-classification performance reaches \( PCC = 0.48 \), suggesting that feature and label heterogeneity across RNA types poses a challenge for unified modeling.

\begin{table}[htb!]
	\small
	\centering
	\caption{Prediction performance of CAP on S142 and S322 datasets.}
	\label{table:PCC2}
	\vspace{8pt}
	\begin{tabular}{l | c | c}
		\hline
		\textbf{Dataset} & \textbf{PCC} & \textbf{RMSE (kcal/mol)} \\
		\hline
		S142  & 0.653 & 2.18 \\
		S322 & 0.669 & 2.00 \\
		\hline
	\end{tabular}
\end{table}

These earlier studies on protein-DNA/RNA affinity predictions have reported results on datasets clustered anywhere from strict 25\% identity to no clustering at all, making headline metrics hard to compare across papers. In our study, we employ the S186 dataset \cite{shen2023svsbi} that solely focus on biochemical uniformity (single protein-nucleic complex per entry, unambiguous labels, identical assay temperature). The resulting dataset, though modest in size, spans a balanced range of binding energies and local neighborhoods, so model performance is driven by genuine physico-chemical signal rather than by the occasional near-duplicate chain. Encouraged by these stable results and to further assess our CAP framework on protein-RNA binding affinity, we curated the S142 dataset of 142 protein-RNA complexes. On this set, our model achieves a \(PCC\) of 0.653 and an RMSE of 2.18 kcal/mol; evaluation on an expanded dataset, S322, of 322 protein-nucleic acid complexes yields better performance (\(PCC\) = 0.669, RMSE = 2.0 kcal/mol). A prediction performance of both datasets can be found in Table~\ref{table:PCC2}, whereas a full comparison of experimental and predicted binding affinities for S142 and S322 are presented in Table~\ref{tab:S142} and Table~\ref{tab:S322}, respectively, along with the corresponding PDBIDs.

Taken together, our CAP framework proposes a more unified approach. It uses a purely sequence-based interaction embedding, built from learned residue and nucleotide descriptors, so it requires neither three-dimensional structures, subclass labels, nor hand-crafted interface features. Furthermore, the existing models in literature run a separate regression model for every DNA/RNA subclass steps that complicate the workflow and risk information loss. Moreover, the datasets used in this study have been carefully curated to include only bona-fide DNA/RNA-binding proteins and complexes containing exactly one protein chain and one DNA/RNA chain. By removing mixed or multifunctional assemblies, the model sees a cleaner and functionally coherent signal with spurious inter-protein contacts that do not contribute to the reported affinity being eliminated.

\section{Datasets }\label{sec:Dataset}

In this section, we give a brief insight about the datasets used in this study, the data collection and curation. Dataset S186 is given by Shen et al. \cite{shen2023svsbi}. It contains mostly protein-DNA complexes but has a few protein-RNA complexes as well. We created a protein-RNA dataset S142 and a protein-nucleic acid dataset S322.  Both datasets were assembled by merging our own curated complexes from PDBbind‐v2020 with some of the complexes used in the PRA-Pred study~\cite{harini2024pred}. We followed the filtering pipeline of Shen \textit{et al.}~\cite{shen2023svsbi}, with several additional steps to improve consistency and reliability. First, all non-standard bases were removed, retaining only \textit{A}, \textit{C}, \textit{G} and \textit{T} for DNA, and \textit{A}, \textit{C}, \textit{G} and \textit{U} for RNA. Complexes having at least one nucleotide were also kept. Although a subset of the S142 and S322 complexes also appears in other studies (e.g., PRdeltaGPred~\cite{hong2023updated} and PredPRBA~\cite{deng2019predprba}), all retained entries passed our independent quality checks. Finally, an additional filter applied to the RNA subsets of both datasets removed systems that do not involve bona-fide RNA-binding proteins, ensuring that the remaining complexes represent genuine RBP-RNA interactions rather than incidental co-crystallisations. RNA-binding proteins often belong to well-studied families such as RNA-recognition motifs (RRMs), K-homology (KH) domains, and double-stranded RNA-binding domains (dsRBDs) \cite{corley2020rna}. These folds rely on recurring chemical strategies: clusters of lysine and arginine neutralise the phosphate backbone, while aromatic side chains stack against exposed bases or the ribose 2$^\prime$-hydroxyl \cite{corley2020rna}. Because those interaction chemistries are reused across many RBPs, the overall contact signatures are comparatively uniform. We therefore restrict our RNA dataset to bona-fide RBPs and group complexes by functional class, giving CAP a chemically coherent training set. DNA-binding proteins, in contrast, encompass a broader spectrum of folds including major-groove readers, backbone clamps, and helix-wrapping architectural proteins and thus, display far more heterogeneous residue-nucleotide contact patterns \cite{yesudhas2017proteins}. The same functional filtering sharpens the RNA dataset but is less beneficial for DNA, where preserving the full diversity of recognition modes is essential. Overall, this unified and rigorously filtered dataset collection underpins the evaluation of CAP. 

\begin{figure}[t!]
	\centering
	\subfigure{
		\includegraphics[width=\textwidth]{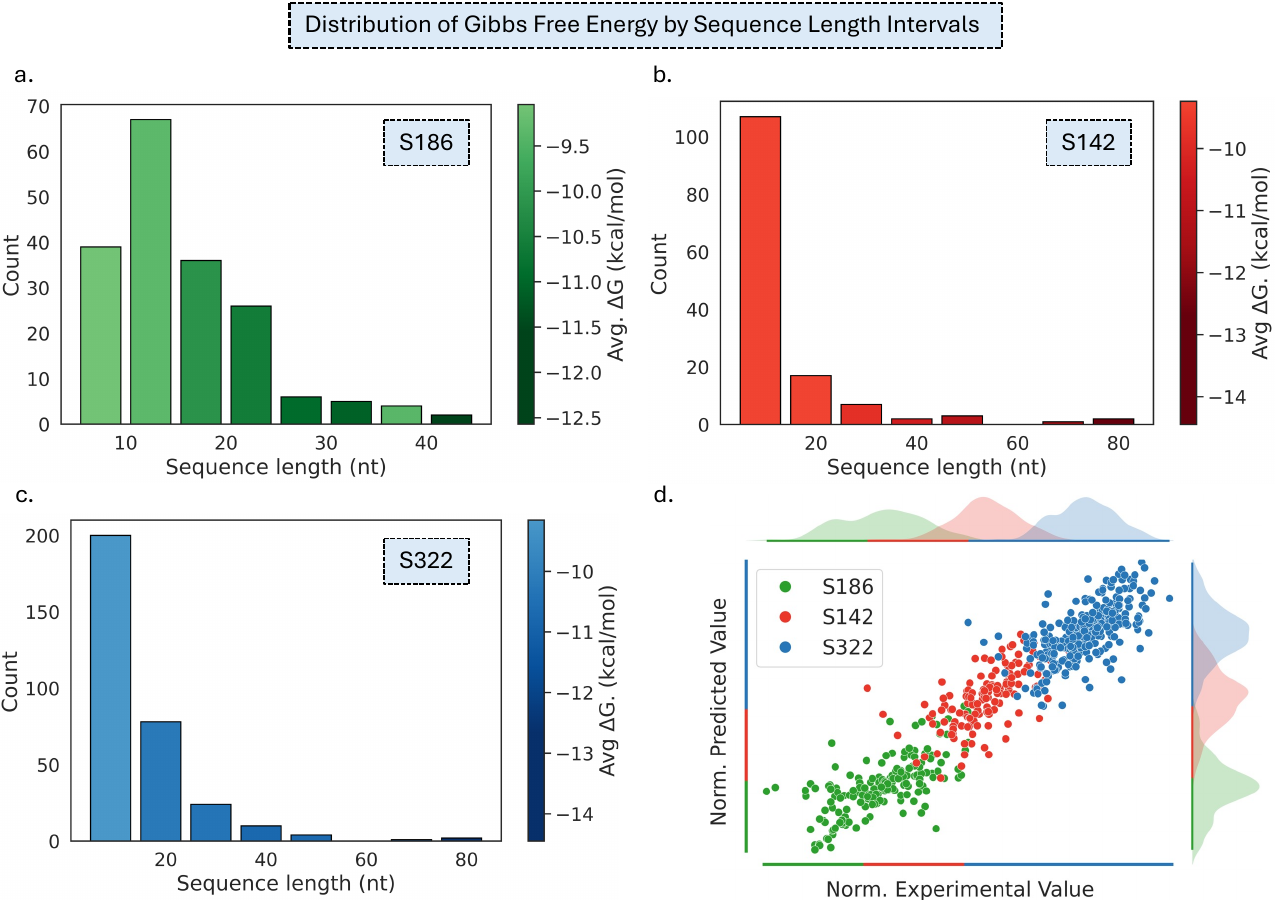}
	}
	\caption{Distribution of experimental Gibbs free energy ($\Delta G$) across sequence length intervals for the datasets (a) S186, (b) S142, and (c) S322. Color intensity encodes mean $\Delta G$; darker shades indicate stronger while lighter shades represents weaker binding. (d) A comparison of normalized experimental and predicted binding affinities of S186, S142, and S322 datasets. Each dataset is scaled to a specific region with an equal range for clear visual illustration.}
	\label{fig:lenDist}
\end{figure}

It is evident that both datasets display a tendency toward more negative (stronger) experimental $\Delta G$ values as sequence length increases, despite sequence counts peaking at shorter lengths. An illustration of how binding free energy varies with sequence length in S142 is represented in Figure~\hyperref[fig:lenDist]{\ref*{fig:lenDist}b}. While the strand lengths range from 4-93$\sim$nt, we note that the mean affinity at the short end (4$\sim$nt), is $-7.05$ kcal/mol, whereas the longest RNAs (93$\sim$nt) bind much more strongly ($\Delta G=-12.27$ kcal/mol). Moreover, we also observe that most RNA complexes (107) fall in the 4-14$\sim$nt interval with a mean $\Delta G=-9.24$ kcal/mol. For the S322 dataset in Figure~\hyperref[fig:lenDist]{\ref*{fig:lenDist}c}, the shortest sequence in the dataset has experimental free-energy change spanning from $-10.56$ to $-4.25$ kcal/mol, with a mean of $-7.05$ kcal/mol. Conversely, the longest sequence in S322 is 93$\sim$nt and shows a single measured value of $\Delta G=-12.27$ kcal/mol. The mean $\Delta G$ for the length interval is $-9.15$ kcal/mol. Figure \hyperref[fig:lenDist]{\ref*{fig:lenDist}d} juxtaposes the experimental and CAP-predicted binding affinities for the S186, S142, and S322 datasets. To aid visual comparison, the points from each dataset are rescaled to the same numerical range and plotted in separate, equally sized bands.

\section{Methods}\label{sec:Methods}

In this section,  we provide an overview of persistent Stanley-Reisner theory (PSRT). Next, we describe the vectorization of persistent commutative algebra along with natural-language processing (NLP) method and model interpretability. Machine learning models and model evaluation metrics are also provided.

\subsection{Persistent Stanley-Reisner Theory}

Persistent Stanley-Reisner theory introduces a new framework for studying data geometry through combinatorial commutative algebra \cite{suwayyid2025persistent}. It maps input data to a simplicial complex made up of vertices, edges, triangles, and cells of higher dimension, thereby preserving both topological and combinatorial characteristics. A filtration records how these characteristics emerge and persist across different spatial scales, yielding algebraic invariants such as persistent facet ideals and adding an algebraic viewpoint to multiscale data analysis. 

Let \(\Delta\) be a simplicial complex on the vertex set \(V = \{x_1, \dots, x_n\}\), and let \((\Delta^t)_{t \in \mathbb{R}}\) be a filtration such that \(\Delta^s \subseteq \Delta^t\) whenever \(s \le t\). Fix a field \(k\), and consider the polynomial ring \(S := k[x_1, \dots, x_n]\). To each \(\Delta^t\), we associate the corresponding {Stanley--Reisner ideal} \(I^t \subseteq S\), defined by
\begin{equation}\label{eq:sr-evolution}
	I^t := I(\Delta^t) := \left\langle x_{i_1} \cdots x_{i_r} \,\middle|\, \{x_{i_1}, \dots, x_{i_r}\} \notin \Delta^t \right\rangle.
\end{equation}
This construction yields a descending sequence of ideals as \(t\) increases.

Each ideal \(I^t\) admits a canonical primary decomposition into monomial prime ideals of the form
\[
I^t = \bigcap_{\sigma \in \mathcal{F}(\Delta^t)} P_\sigma, \quad \text{where} \quad P_\sigma := (x_i \mid x_i \notin \sigma) \subseteq S,
\]
and \(\mathcal{F}(\Delta^t)\) denotes the set of facets of \(\Delta^t\). The ideals \(P_\sigma\) are referred to as the {facet ideals} at level \(t\).

Let
\[
\mathcal{P}^t := \{ P_\sigma \mid \sigma \in \mathcal{F}(\Delta^t) \}
\]
denote the set of facet ideals associated to \(\Delta^t\). To analyze the structure of \(\mathcal{P}^t\) by dimension, for each \(i \ge 0\), define the graded stratification
\[
\mathcal{P}_i^t := \left\{ P_\sigma \in \mathcal{P}^t \,\middle|\, \dim(\sigma) = i \right\},
\]
so that
\[
\mathcal{P}^t = \bigsqcup_{i = 0}^{\dim(\Delta^t)} \mathcal{P}_i^t.
\]

In analogy with persistent homology, we define the notion of {persistent facet ideals} and their associated facet persistent numbers. A facet ideal \(P_\sigma \in \mathcal{P}_i^t\) is said to {persist} to level \(t' > t\) if \(P_\sigma \in \mathcal{P}_i^{t'}\). The set of such persistent \(i\)-dimensional facet ideals is denoted by
\[
\mathcal{P}_i^{t,t'} := \mathcal{P}_i^t \cap \mathcal{P}_i^{t'}.
\]

The corresponding {facet persistent number} is defined as
\[
\mathcal{F}_i^{t,t'} := \left| \mathcal{P}_i^{t,t'} \right|,
\]
which counts the number of \(i\)-dimensional facet ideals common to both levels \(t\) and \(t'\).

The collection \(\{ \mathcal{F}_i^{t,t'} \}_{i,\,t,\,t'}\) defines a graded combinatorial invariant that captures the evolution of the minimal prime decompositions of the Stanley--Reisner ideals across the filtration. These invariants quantify the persistence of \(i\)-dimensional facet ideals and may be regarded as an algebraic counterpart to topological barcodes in persistent homology.

In the special case where \(t = t'\), the function
\[
t \mapsto \mathcal{F}_i^{t,t}
\]
is referred to as the {facet curve} of degree \(i\). It records the number of \(i\)-dimensional facet ideals appearing in the minimal prime decomposition of \(I^t\) at each level \(t\).

Furthermore, the normalized quantity
\[
\frac{\mathcal{F}_i^{t,t}}{t}
\]
is called the {persistence rate} of the \(i\)th facet persistent number at time \(t\), and serves as a measure of the density of \(i\)-dimensional prime components within the decomposition of \(I^t\).

\subsection{1-mer Algebra}

In this section, we present the specialization of the \(k\)-mer algebra framework to the case \(k=1\). 
This method derives an algebraic structure through the Stanley-Reisner construction, based on the positional distribution of individual nucleotides.
Let \(\mathcal{A}\) be \(\{A,\, C,\, G,\, T/U\}\), and let
\[
S = s_1 s_2 \cdots s_N \;\in\; \mathcal{A}^N
\]
be a sequence of length \(N\).  For each letter \(a \in \mathcal{A}\), define the {indicator function}
\[
\delta_a \colon \mathcal{A} \to \{0,1\}, 
\qquad
\delta_a(b) =
\begin{cases}
	1, & b = a,\\
	0, & b \neq a.
\end{cases}
\]

\medskip

The set of positions at which \(a\) occurs in \(S\) is
\[
S^a \;=\;
\bigl\{\, i \in \{1,\dots,N\} \;\bigm|\;\delta_a(s_i)=1 \bigr\}
\;\subset\; \mathbb{N}.
\]
We view \(S^a\subset \mathbb{R}\) as a one-dimensional input data.  Define the corresponding {pairwise distance matrix}
\[
D^a = \bigl(d^a_{ij}\bigr)_{i,j \in S^a}
\in \mathbb{R}^{|S^a|\times |S^a|},
\qquad
d^a_{ij} = \lvert i - j\rvert.
\]

\medskip

Fix a filtration interval \([t_0,t_1]\subset \mathbb{R}_{\ge0}\).  For each \(t\in[t_0,t_1]\), construct the Vietoris–Rips complex on \(S^a\) at scale \(t\) and compute its algebraic invariants.  Denote by
\[
v^t_a = \bigl(v^t_i(a)\bigr)_{i\in\mathbb{N}}
\]
the vector of algebraic invariants in each dimension \(i\in\mathbb{N}\) (e.g.\ facet vectors $\mathcal{F}$, and facet rate vectors.). Restricting to a fixed scale \(t\), we write
\[
v_a := \bigl(v_i(a)\bigr)_{i\in\mathbb{N}}.
\]

\medskip

Finally, the full 1-mer representation of \(S\) is obtained by concatenation over all \(a\in\mathcal{A}\):
\[
\boldsymbol{v}^1_S 
:= \bigl(v_a\;\bigm|\;a\in\mathcal{A}\bigr)
= \bigl(\boldsymbol{v}^1_{S, i})_{i\in\mathbb{N}},
\]
where \( \boldsymbol{v}^1_{S, i} := \left( v_i(A),\, v_i(C),\, v_i(G),\, v_i(T/U) \right) \) is the vector of dimension-\( i \) features computed over all \( 1 \)-mers in \( S \). 
We refer to \(\boldsymbol{v}^1_S\) as the CAP-vector representation of \(S\) at \(k=1\), which encodes the spatial distribution of each symbol via persistent algebraic invariants.

\subsection{Vectorization of persistent comutative algebra}\label{sec:Vectorization}

To capture the intrinsic structure of nucleic acid sequence, we encode every sequence as a one-dimensional input data, where each point indicates the position of a mononucleotide (i.e., $\textit{k}$-mers of length 1). To ensure consistency, we applied a uniform grid of fifty equally spaced filtration thresholds, \(\;r = 0,1,2,\dots,49\;\), to every \(k\)-mer input. A Vietoris-Rips filtration is then applied using these fixed radii across all samples. This setup generates uniformly aligned facet-count curves and their corresponding rate profiles, each sampled at fifty evenly spaced filtration values. As a result, we derive two parallel families of features with a filtration design that captures structural regularities while suppressing noise and guarding against outliers.

For each sequence, we identify the positions of all four mononucleotides (A, C, G, T/U), forming four separate inputs. To each input, we apply persistent commutative algebra modeling, using persistent facet ideals to measure the algebraic complexity of interaction structures across scales. For every mononucleotide-specific cloud, we compute 50-dimensional facet curves in dimensions 0, 1, and 2, along with 50-dimensional rate curves capturing the change in facet growth across the filtration. This yields 150 features from the facet counts and 150 from the facet rates, resulting in 300 features per nucleotide. Concatenating the features across all four nucleotide types produces a 1200-dimensional feature vector that summarizes sequence-derived structural patterns. This coherent representation of simplicial structure and connectivity evolution across an identical filtration range yields a rich input for our binding-affinity model. 

\subsection{Transformer-based protein language model}\label{sec:NLP}

Recent advances in natural-language processing (NLP) provide powerful sequence-based insights for molecular biosciences. We harness these techniques to augment our PSRT framework for predicting protein-nucleic acid binding affinities. While PSRT analyzes nucleic acid sequences, NLP captures patterns  directly from amino acid sequences.

To complement the commutative-algebraic descriptors of DNA/RNA, we incorporate protein sequence information using a state-of-the-art transformer-based language model, ESM2 \cite{lin2023evolutionary}. This model processes the raw amino acid sequence and outputs 2560-dimensional high-level embeddings by aggregating across 36 transformer layers. The final feature matrix for each protein-nucleic acid complex is constructed by concatenating the 1200-dimensional PCA-derived nucleotide features with the 2560-dimensional ESM2-based protein features, resulting in a unified 3760-dimensional representation. These composite features are used to train and evaluate CAP on our dataset. An illustration of the workflow in CAP is demonstrated in \autoref{fig:concept}.

\subsection{Machine learning modeling}\label{sec:ML}

We train our regression models with Gradient Boosting Decision Trees (GBDT), implemented in Python via the \texttt{scikit-learn} library\,(v1.3.2). GBDT is valued for its resistance to overfitting, limited sensitivity to hyperparameter tuning, and straightforward deployment. It assembles many shallow decision trees generated from bootstrap samples of the training data and aggregates their outputs, so the ensemble corrects errors that any single tree might make. We supply the algorithm separately with PSRT-derived molecular descriptors and transformer-based descriptors; the hyperparameter settings are listed in \autoref{table:GBDT-parameters}.

\begin{table}[htb!]
	\small
	\centering
		\caption{Hyperparameters used in the gradient boosting regression model within a scikit-learn pipeline. StandardScaler is applied prior to model training.}
	\begin{tabular}{c c c c}
		\hline
		\textbf{No. of estimators} & \textbf{Max depth} & \textbf{Min. samples split} & \textbf{Learning rate} \\
		10,000 & 7 & 3 & 0.01 \\
		\hline
		\textbf{Max features} & \textbf{Subsample size} & \textbf{Random state} & \textbf{Standardization} \\
		Square root & 0.7 & Fixed (seeded) & Yes (StandardScaler) \\
		\hline
	\end{tabular}
	\label{table:GBDT-parameters}
\end{table}

\subsection{Evaluation metrics}

To quantitatively evaluate the performance of our binding affinity prediction models, we employ the Pearson correlation coefficient (\(PCC\)), defined as:
\begin{align*}
	\text{PCC}(\mathbf{x}, \mathbf{y}) = \frac{\sum_{m=1}^{M} (y_m^e - \bar{y}^e)(y_m^p - \bar{y}^p)}{\sqrt{\sum_{m=1}^{M} (y_m^e - \bar{y}^e)^2 \sum_{m=1}^{M} (y_m^p - \bar{y}^p)^2}},
\end{align*}
where \( y_m^e \) and \( y_m^p \) denote the experimental and predicted binding affinity values for the \( m \)-th sample, respectively, and \( \bar{y}^e \) and \( \bar{y}^p \) are their corresponding mean values.

We also report the root mean squared error (RMSE), which is computed as:
\begin{align*}
	\mathrm{RMSE} = \sqrt{\frac{1}{n} \sum_{m=1}^{M} (y_m^e - y_m^p)^2},
\end{align*}
where \( y_m^e \) and \( y_m^p \) represent the experimental and predicted binding affinity values for the \( m \)-th sample, respectively.

The above two metrics are employed to assess the performance of our machine learning models on all datasets. The original labels for these datasets are given as \( pK_d \) values, which are converted to binding free energies (in kcal/mol) by multiplying by a constant factor of 1.3633. Our models achieve reasonable RMSE values across all three datasets.

 \section{Model interpretability}\label{sec:Interpretability}

\begin{figure}[t!]
	\centering
	\subfigure{
		\includegraphics[width=\textwidth]{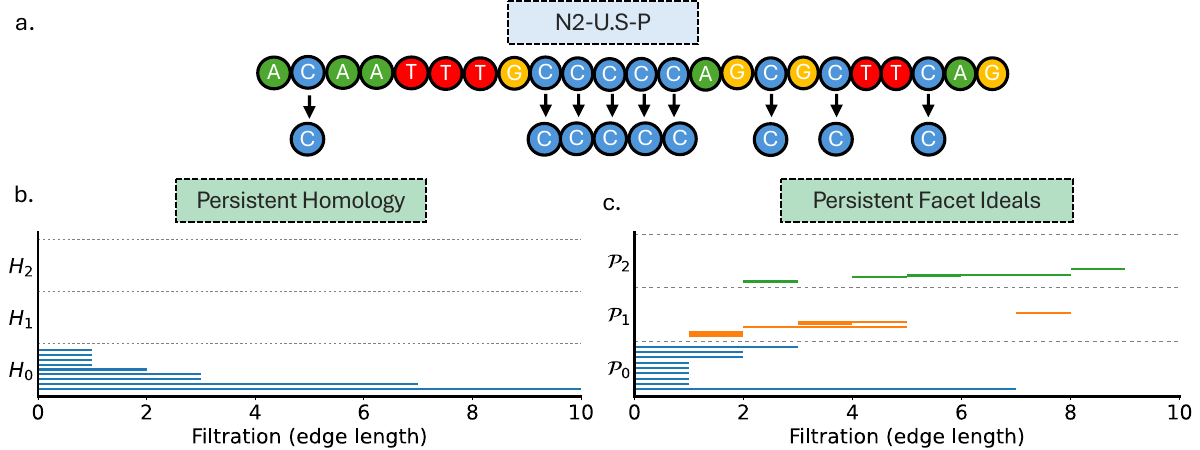}
	}
		\caption{  Comparison between Persistent Homology and Persistent Stanley-Reisner Invariants.  (a) The N2-U.S-P primer sequence, with the positions of the nucleotide C marked. (b) The persistent homology barcodes computed from the input set formed by the positions of C, representing topological features in the spatial distribution. (c) The persistent facet barcodes derived of the same input set, encoding combinatorial face-level patterns rather than topological invariants and reflecting the algebraic invariants of maximal simplices under the same induced filtration.}
	\label{fig:covid}
\end{figure}

Topological techniques like persistent homology (PH) \cite{zomorodian2004computing} and persistent Laplacian theory \cite{wang2020persistent} have emerged as powerful tools in machine learning frameworks. Although our persistent commutative algebra technique is built on the same simplicial‐complex scaffold of PH however, it unravels the combinatorial activity by shifting the focus from a global view to a fine-grained, local view across the same filtration. Moreover, in comparison with persistent Laplacian theory \cite{wang2020persistent}, which can be computationally intensive in terms of analyzing eigenvalues of large Laplacian matrices, our algebraic construction sidesteps matrix diagonalization and scales more efficiently. Each of these  varying paradigms offers its own unique advantages, making them suitable for different data regimes and analytical goals.

In this section, we demonstrate how a simplicial complex evolves distinctively under a growing filtration parameter for PH and persistent Stanley-Reisner (SR) approach. For this purpose, we analyze the N2-U-S-P primer sequence as shown in Figure~\hyperref[fig:covid]{\ref*{fig:covid}a}. The 23-nucleotide primer contains several cytosines that we have highlighted; only these C residues are used to build the one-dimensional input set in this example, placing each point at its sequence index (positions 2, 9, 10, 11, 12, 13, 16, 18, and 21). This resulting input set is analyzed using both persistent homology and persistent facet ideals.

Figure~\hyperref[fig:covid]{\ref*{fig:covid}b} shows the barcodes from the persistent homology analysis, which records the appearance and disappearance of global features - connected components, loops, and voids - as the Vietoris–Rips filtration radius \(r\) increases. For this cytosine-based one-dimensional input, the $H_0$ barcode comprises blue bars, each initially representing an isolated cytosine at $r=0$. As the filtration expands, these points progressively merge, reducing the number of separate clusters. Notably, no bars appear in the $H_1$ or $H_2$ lanes, indicating that higher-dimensional features such as loops and cavities are not formed in this case. On the contrary, Figure~\hyperref[fig:covid]{\ref*{fig:covid}c} shows barcodes derived from the persistent facet ideals applied to the same filtration data. The zeroth-dimensional panel $P_0$ barcodes (blue bars) shows the appearance of isolated components. The edges appear around $r = 1$  in the first-dimensional panel $P_1$ (orange bars) which was entirely missed by persistent homology. Similarly, the second-dimensional panel $P_2$ (green bars) captures the birth and death of 2-simplices around $r = 2$, indicating the formation of triangular faces in the structure as the filtration progresses.

Persistent homology and persistent facet ideals both analyze how a simplicial complex evolves under a growing filtration parameter, but they focus on different features. Unlike persistent homology which is summarized by the Betti‐number barcodes, persistent facet ideals directly capture local simplex activities (vertices, edges, triangles, and higher-order structures) throughout the filtration process. The filtration process involves gradually increasing a radius parameter $r$ to monitor the formation and dissolution of topological features encoded via facet ideals. For example, the $H_{0}$ barcode records the number of connected components: it begins at the total number of vertices and decreases by one each time an edge appears, reaching a single component when the complex becomes fully connected. In contrast, the facet‐ideal count $P_{0}$ measures the number of isolated vertices (0-simplices not yet incident to any edge) and thus falls to zero as soon as every vertex participates in at least one edge. This facet-based barcode framework captures both geometric and combinatorial properties of molecules. Thus, persistent homology emphasizes the emergence of homology classes that change the global topology, while persistent facet ideals specialize in capturing precise local combinatorial events across the same filtration. This enhances interpretability and relevance for biomolecular sequence analyses, thereby improving subsequent performance in machine learning applications.

\section{Conclusion}

Commutative algebra, traditionally central to fields like algebraic geometry and number theory, has remained largely untapped in data-driven and biological research. Recent work by Suwayyid and Wei \cite{suwayyid2025persistent} changes this by integrating Stanley-Reisner theory with multiscale analysis, opening a new avenue for nonlinear algebraic techniques in data science. Initial studies, such as those on protein-ligand binding affinity \cite{feng2025caml}, attest to the power of the persistent Stanley-Reisner theory (PSRT) framework for enhancing machine learning models.

In this work, we present commutative algebra prediction (CAP), a framework for protein-nucleic acid binding affinity prediction that integrates PSRT-based $\textit{k}$-mer algebra descriptors of DNA sequences and transformer-derived ESM2 embeddings of protein sequences. We evaluate the performance of our model on the S186 dataset with an additional test on newly curated datasets S142 and S322. CAP is trained with a single gradient-boosting regression tree model to provide variance across splits without the need for iterative feature pruning or subclass-specific retraining. Our findings deliver strong and generalizable predictions with Pearson correlation coefficient of 0.705 with RMSE of 1.79 kcal/mol outperforming SVSBI \cite{shen2023svsbi} (0.669/1.98 kcal/mol). Furthermore, CAP attains a \(PCC\) of 0.653 and an RMSE of 2.18 kcal/mol on S142, while its performance improves on S322, reaching a \(PCC\) of 0.669 with an RMSE of 2.0 kcal/mol. These results demonstrate the robustness of CAP in jointly capturing informative DNA k-mer algebra signatures and protein ESM2 embeddings, thereby enabling reliable prediction of protein-nucleic acid binding affinities. Finally, because CAP’s inputs are limited to primary sequences, the framework is inherently scalable: it can be applied to any protein-DNA/RNA pair with minimal preprocessing and is readily extensible to genome-scale studies. In addition to this, the proposed commutative algebraic methodology can not only be applied to the study of other biological sequences, such as phylogenetic analysis and protein sequence representation but it can also be generalized to the analysis of other sequential data in science and engineering.    


\section*{Data and Code availability}
All codes and datasets needed to evaluate the conclusions in this study are available at \href{https://github.com/mzia-s/CAP}{https://github.com/mzia-s/CAP}. The SVSBI dataset can also be accessed at \href{https://github.com/WeilabMSU/SVS}{https://github.com/WeilabMSU/SVS}.

\section*{Supporting Information}
\href{./SI.pdf}{Supplementary Information} is available for supplementary tables.

\section*{Conflict of Interest}

The authors declare no competing financial interests.

\section*{Acknowledgments}
This work was supported in part by NIH grant R35GM148196, NSF grant DMS-2052983, and MSU Research Foundation. F.S. thanks King Fahd University of Petroleum and Minerals for their support.

\clearpage

\bibliographystyle{unsrt}
\bibliography{refs}

\end{document}


\maketitle 
\newpage	
	{\setcounter{tocdepth}{4} \tableofcontents}
	\setcounter{page}{1}
	\newpage	
	
	\clearpage

	\section{Results: Experimental vs. Predicted Binding Affinities}
	\vspace{1em}
	
	\subsection{The S186 Dataset}

	\renewcommand*{\arraystretch}{1.2}
	\setlength{\tabcolsep}{6pt}
	
{\centering
\begin{longtable}{%
		>{\centering\arraybackslash}p{1.5cm}
		>{\centering\arraybackslash}p{1.5cm}
		>{\centering\arraybackslash}p{1.5cm}
		>{\centering\arraybackslash}p{1.5cm}
		>{\centering\arraybackslash}p{1.5cm}
		>{\centering\arraybackslash}p{1.5cm}
		>{\centering\arraybackslash}p{1.5cm}
		>{\centering\arraybackslash}p{1.5cm}
		>{\centering\arraybackslash}p{1.5cm}%
	}
	
\caption[Binding free energies for the PN dataset]%
{Experimental and predicted binding free energies ($\Delta G$) for the S186 dataset \cite{shen2023svsbi}.\label{tab:S186}}\\

\toprule
\textbf{PDBID} & \textbf{Exp BA} & \textbf{Pred BA} &
\textbf{PDBID} & \textbf{Exp BA} & \textbf{Pred BA} &
\textbf{PDBID} & \textbf{Exp BA} & \textbf{Pred BA}\\
\midrule
\endfirsthead

\toprule
\textbf{PDBID} & \textbf{Exp BA} & \textbf{Pred BA} &
\textbf{PDBID} & \textbf{Exp BA} & \textbf{Pred BA} &
\textbf{PDBID} & \textbf{Exp BA} & \textbf{Pred BA}\\
\midrule
\endhead

			1HVO & -6.760 & -7.227 & 1WET & -7.940 & -11.516 & 2PUF & -7.769 & -11.158 \\
			1QPZ & -11.704 & -10.246 & 1QP0 & -10.165 & -10.955 & 1QP7 & -8.327 & -8.392 \\
			1FJX & -12.070 & -12.671 & 1J75 & -10.165 & -9.392 & 1JFS & -11.803 & -10.957 \\
			1J5K & -7.529 & -9.805 & 1PO6 & -9.276 & -9.696 & 1P51 & -11.528 & -11.390 \\
			1QZG & -8.614 & -9.464 & 1OSB & -9.754 & -9.754 & 1HVN & -7.227 & -6.758 \\
			1BDH & -8.242 & -8.118 & 1QP4 & -10.906 & -11.221 & 1QQB & -7.769 & -8.664 \\
			1TW8 & -12.192 & -12.234 & 1QQA & -8.722 & -8.250 & 1DH3 & -12.070 & -10.448 \\
			1JJ4 & -11.991 & -10.913 & 1FYM & -10.086 & -9.559 & 1JH9 & -11.859 & -7.913 \\
			1JT0 & -9.965 & -10.387 & 1P78 & -11.528 & -11.555 & 1P71 & -11.528 & -11.296 \\
			1OMH & -9.754 & -9.754 & 1S40 & -12.983 & -10.193 & 1U1L & -9.503 & -8.938 \\
			1U1R & -9.053 & -9.165 & 1U1N & -8.862 & -9.677 & 1U1O & -9.855 & -9.646 \\
			2BJC & -14.996 & -10.129 & 1ZZI & -7.713 & -9.770 & 2B0D & -8.242 & -9.134 \\
			2GII & -12.783 & -12.781 & 2GIE & -12.592 & -12.582 & 2GIH & -12.252 & -12.254 \\
			2ADY & -10.707 & -10.494 & 2AC0 & -10.364 & -10.013 & 2AYB & -11.859 & -11.797 \\
			2ERE & -10.256 & -10.295 & 2GE5 & -8.751 & -9.147 & 2O6M & -10.751 & -10.058 \\
			3D6Z & -8.204 & -6.798 & 2VYE & -9.991 & -9.671 & 3IGM & -8.590 & -9.391 \\
			3HXQ & -12.572 & -12.566 & 3EQT & -9.514 & -9.864 & 3H15 & -7.455 & -8.684 \\
			2KAE & -11.039 & -9.858 & 3MQ6 & -12.572 & -10.865 & 3JSP & -11.995 & -12.031 \\
			3AAF & -8.994 & -9.236 & 3N1L & -11.407 & -11.239 & 3N1K & -11.236 & -11.406 \\
			3R8F & -9.487 & -9.775 & 2RRA & -7.940 & -9.453 & 3QMI & -6.709 & -6.639 \\
			2KXN & -7.700 & -9.556 & 1U1P & -9.107 & -9.274 & 1U1M & -8.901 & -9.165 \\
			1U1K & -8.856 & -9.141 & 1U1Q & -9.706 & -9.586 & 2AOR & -8.242 & -8.299 \\
			2AOQ & -7.438 & -8.094 & 2I9K & -12.983 & -11.832 & 2GIJ & -12.783 & -12.781 \\
			2GIG & -12.252 & -12.254 & 2CCZ & -9.543 & -10.346 & 2AHI & -10.496 & -10.705 \\
			2ERG & -10.364 & -10.034 & 2AYG & -11.859 & -11.765 & 2ES2 & -8.843 & -9.358 \\
			2ATA & -9.646 & -10.400 & 2NP2 & -9.897 & -10.567 & 3D6Y & -6.799 & -8.206 \\
			2VY1 & -9.573 & -9.888 & 3HXO & -12.572 & -12.572 & 3GIB & -12.070 & -9.482 \\
			3D2W & -9.355 & -9.032 & 2KKF & -5.939 & -7.859 & 3M9E & -9.388 & -9.799 \\
			3JSO & -12.402 & -11.641 & 3K3R & -11.966 & -12.244 & 3N1I & -11.554 & -10.859 \\
			3N1J & -11.386 & -10.859 & 3KJP & -11.180 & -9.631 & 3Q0B & -8.134 & -9.231 \\
			3PIH & -11.890 & -10.408 & 3QMH & -6.502 & -6.823 & 3QMB & -7.438 & -6.574 \\
			3U7F & -8.482 & -9.231 & 3RN2 & -9.476 & -9.011 & 4ATK & -11.956 & -11.850 \\
			2LTT & -10.628 & -10.156 & 4HIO & -10.467 & -10.106 & 4HIK & -10.388 & -10.374 \\
			4A76 & -9.855 & -9.274 & 4HJ7 & -11.209 & -10.445 & 4HID & -8.273 & -10.424 \\
			4LJ0 & -8.590 & -9.187 & 4HT8 & -9.471 & -11.451 & 4GCK & -9.256 & -9.742 \\
			4F2J & -10.364 & -10.110 & 4GCT & -9.915 & -9.872 & 4NI7 & -13.223 & -9.962 \\
			4CH1 & -9.003 & -9.149 & 4QJU & -8.768 & -10.589 & 4ZBN & -13.194 & -10.142 \\
			4R55 & -7.876 & -9.658 & 4S0N & -10.751 & -9.311 & 4ZSF & -11.639 & -10.816 \\
			4RKG & -6.258 & -9.099 & 5A72 & -9.916 & -10.327 & 5DWA & -12.332 & -10.915 \\
			5K83 & -7.913 & -6.797 & 5ITH & -10.256 & -9.518 & 5W9S & -9.543 & -8.161 \\
			5YI3 & -8.534 & -9.730 & 5XFP & -8.072 & -9.584 & 5MEY & -9.265 & -9.084 \\
			5MEZ & -9.001 & -9.356 & 5W2M & -7.084 & -7.453 & 5K17 & -6.891 & -9.412 \\
			6MG1 & -10.276 & -11.106 & 5VMV & -12.769 & -10.045 & 6CNQ & -8.242 & -9.063 \\
			6FQP & -9.229 & -9.342 & 5WWF & -9.029 & -9.040 & 6FQQ & -8.739 & -9.349 \\
			5ZVB & -5.990 & -5.940 & 5ZVA & -5.809 & -6.266 & 5MPF & -9.605 & -9.887 \\
			6G1L & -12.332 & -11.459 & 6IIQ & -8.024 & -9.616 & 6IIR & -7.570 & -9.127 \\
			3ON0 & -11.341 & -10.466 & 4HQU & -14.586 & -11.900 & 4HQX & -12.162 & -13.442 \\
			4A75 & -9.232 & -9.818 & 3QSU & -10.511 & -10.019 & 4HJ8 & -10.440 & -8.511 \\
			4HIM & -10.132 & -10.470 & 4HJ5 & -9.817 & -10.383 & 4HP1 & -7.028 & -8.751 \\
			4J1J & -8.584 & -10.111 & 4NM6 & -8.140 & -9.119 & 4GCL & -9.278 & -9.826 \\
			3ZPL & -11.751 & -10.355 & 3ZH2 & -10.057 & -10.866 & 4HT4 & -11.039 & -9.790 \\
			4LNQ & -8.011 & -9.152 & 4LJR & -10.244 & -10.216 & 4R56 & -9.370 & -6.700 \\
			4TMU & -12.030 & -7.978 & 4R22 & -10.819 & -10.223 & 4Z3C & -7.661 & -8.187 \\
			3WPC & -10.496 & -11.437 & 3WPD & -11.619 & -10.389 & 2N8A & -9.573 & -9.650 \\
			5HRT & -12.114 & -10.186 & 5T1J & -10.526 & -9.908 & 5VC9 & -9.265 & -7.710 \\
			5HLG & -8.873 & -10.191 & 5W9Q & -8.482 & -7.733 & 6ASB & -7.940 & -9.008 \\
			6ASD & -7.661 & -8.559 & 5K07 & -6.122 & -9.359 & 5YI2 & -9.738 & -8.594 \\
			6MG3 & -11.449 & -10.330 & 6FWR & -8.873 & -10.244 & 6CNP & -8.391 & -8.553 \\
			5ZD4 & -10.798 & -9.562 & 5ZMO & -9.163 & -9.907 & 6BWY & -6.251 & -9.430 \\
			6CRM & -7.264 & -11.533 & 6CC8 & -7.181 & -9.183 & 6BUX & -5.807 & -8.405 \\
			6A2I & -8.701 & -7.031 & 5ZKI & -8.015 & -10.158 & 6KBS & -7.713 & -9.911 \\
			5ZKL & -10.479 & -10.177 & 5ZMD & -7.407 & -9.448 & 6ON0 & -9.605 & -9.688 \\

\bottomrule
\end{longtable}
}
	
	\newpage	
\subsection{The S142 Dataset}

\renewcommand*{\arraystretch}{1.2}
\setlength{\tabcolsep}{6pt}

{\centering
	\begin{longtable}{%
			>{\centering\arraybackslash}p{1.5cm}
			>{\centering\arraybackslash}p{1.5cm}
			>{\centering\arraybackslash}p{1.5cm}
			>{\centering\arraybackslash}p{1.5cm}
			>{\centering\arraybackslash}p{1.5cm}
			>{\centering\arraybackslash}p{1.5cm}
			>{\centering\arraybackslash}p{1.5cm}
			>{\centering\arraybackslash}p{1.5cm}
			>{\centering\arraybackslash}p{1.5cm}%
		}
	
\caption[Binding free energies for the PN dataset]%
{Experimental and predicted binding free energies ($\Delta G$) for the S142 dataset.\label{tab:S142}}\\
	
	\toprule
	\textbf{PDBID} & \textbf{Exp BA} & \textbf{Pred BA} &
	\textbf{PDBID} & \textbf{Exp BA} & \textbf{Pred BA} &
	\textbf{PDBID} & \textbf{Exp BA} & \textbf{Pred BA}\\
	\midrule
	\endfirsthead
	
	\toprule
	\textbf{PDBID} & \textbf{Exp BA} & \textbf{Pred BA} &
	\textbf{PDBID} & \textbf{Exp BA} & \textbf{Pred BA} &
	\textbf{PDBID} & \textbf{Exp BA} & \textbf{Pred BA}\\
	\midrule
	\endhead

	2A9X & -9.303 & -10.928 & 3BX3 & -9.874 & -10.332 & 3IRW & -14.996 & -11.417 \\
	3IWN & -12.270 & -13.614 & 3K0J & -10.992 & -13.009 & 3QGB & -11.528 & -11.324 \\
	2XS2 & -10.113 & -7.450 & 3QGC & -10.982 & -10.932 & 4GHA & -8.590 & -9.752 \\
	3V6Y & -11.464 & -11.571 & 2LUP & -6.090 & -10.548 & 4M59 & -9.531 & -12.045 \\
	2MTV & -9.526 & -7.434 & 4RCM & -6.350 & -8.934 & 5EIM & -8.123 & -8.661 \\
	5GXH & -7.661 & -9.294 & 5KLA & -10.334 & -11.841 & 5EN1 & -8.942 & -8.889 \\
	5F5H & -9.668 & -9.571 & 5WZK & -9.265 & -10.337 & 5HO4 & -9.462 & -9.306 \\
	5U9B & -9.388 & -8.770 & 5WZJ & -10.751 & -10.317 & 5TF6 & -11.639 & -10.081 \\
	5SZE & -10.459 & -8.724 & 5WZG & -10.218 & -10.508 & 5M8I & -5.736 & -8.548 \\
	5UDZ & -9.592 & -9.059 & 5YTV & -8.043 & -7.852 & 6FQ3 & -8.015 & -9.048 \\
	5YTX & -8.007 & -8.022 & 5YKI & -12.030 & -9.858 & 6DCL & -10.647 & -9.134 \\
	6GD3 & -5.191 & -6.956 & 6G2K & -6.427 & -6.652 & 6CMN & -11.727 & -9.367 \\
	5YTS & -7.577 & -7.648 & 5WWG & -9.077 & -8.823 & 5WWF & -9.029 & -8.892 \\
	5YTT & -7.394 & -7.691 & 6FQR & -6.073 & -7.340 & 6GX6 & -7.264 & -6.194 \\
	5WWE & -8.550 & -9.024 & 6GC5 & -8.435 & -6.979 & 6RT6 & -6.755 & -7.770 \\
	6NOF & -10.617 & -10.477 & 6R7B & -10.558 & -9.084 & 6NOC & -10.490 & -10.527 \\
	6NOH & -10.421 & -10.602 & 6A6J & -9.068 & -8.372 & 6NOD & -10.280 & -11.359 \\
	6G99 & -6.176 & -9.008 & 6RT7 & -7.405 & -7.234 & 6U9X & -10.086 & -9.895 \\
	6NY5 & -9.952 & -10.127 & 6GBM & -5.556 & -9.515 & 1EC6 & -7.990 & -8.866 \\
	1M8Y & -9.270 & -12.410 & 1RPU & -13.300 & -9.820 & 1UTD & -16.890 & -8.985 \\
	2B6G & -10.600 & -9.754 & 2ERR & -12.200 & -8.668 & 2F8K & -10.300 & -9.725 \\
	2G4B & -7.830 & -7.809 & 2KFY & -8.840 & -8.056 & 2KG0 & -7.400 & -8.786 \\
	2KG1 & -7.450 & -8.247 & 2KX5 & -11.400 & -10.733 & 2KXN & -8.170 & -8.775 \\
	2L41 & -4.250 & -7.057 & 2LA5 & -11.500 & -10.957 & 2LEB & -9.320 & -9.455 \\
	2LEC & -9.440 & -9.339 & 2M8D & -8.730 & -8.857 & 2MJH & -9.670 & -9.264 \\
	2MXY & -7.900 & -6.836 & 2MZ1 & -6.870 & -7.886 & 2N82 & -10.300 & -9.827 \\
	2RRA & -7.940 & -8.220 & 2RU3 & -8.520 & -9.256 & 2XC7 & -7.040 & -8.121 \\
	2XFM & -8.240 & -7.664 & 2XNR & -5.400 & -7.537 & 2ZKO & -8.050 & -10.836 \\
	3BSB & -10.200 & -10.686 & 3BSX & -11.500 & -11.098 & 3BX2 & -9.970 & -9.783 \\
	3EQT & -9.520 & -9.285 & 3GIB & -10.690 & -9.402 & 3K49 & -12.050 & -10.730 \\
	3K4E & -10.800 & -11.844 & 3K5Q & -9.300 & -9.172 & 3K5Y & -10.540 & -9.425 \\
	3K5Z & -9.380 & -9.951 & 3K61 & -8.890 & -9.677 & 3K62 & -8.740 & -9.250 \\
	3K64 & -9.220 & -8.941 & 3L25 & -8.175 & -9.291 & 3LQX & -12.220 & -10.107 \\
	3MDG & -7.850 & -8.737 & 3MOJ & -13.000 & -10.506 & 3NCU & -10.300 & -9.229 \\
	3NMR & -8.440 & -7.356 & 3NNH & -6.190 & -8.264 & 3O3I & -6.520 & -7.069 \\
	3O6E & -7.070 & -6.522 & 3Q0L & -12.300 & -11.270 & 3Q0M & -12.500 & -12.366 \\
	3Q0N & -10.500 & -10.977 & 3Q0P & -12.800 & -10.299 & 3Q0Q & -13.300 & -12.526 \\
	3Q0R & -13.800 & -12.844 & 3Q0S & -10.800 & -12.416 & 3QG9 & -10.600 & -10.987 \\
	3U4M & -15.900 & -10.326 & 3V71 & -11.100 & -10.255 & 3V74 & -11.600 & -11.382 \\
	3WBM & -9.980 & -10.926 & 4CIO & -9.760 & -8.359 & 4ED5 & -9.080 & -8.344 \\
	4ERD & -9.540 & -9.409 & 4HT8 & -9.380 & -10.205 & 4JVH & -9.590 & -9.179 \\
	4KJI & -8.940 & -10.087 & 4LG2 & -9.540 & -9.117 & 4NL3 & -7.380 & -10.254 \\
	4O26 & -8.200 & -11.597 & 4OE1 & -12.110 & -9.503 & 4QI2 & -9.440 & -9.509 \\
	4QVC & -11.560 & -10.060 & 4QVD & -10.120 & -10.314 & 4R3I & -7.770 & -6.755 \\
	4RCJ & -8.180 & -8.213 & 4TUX & -10.400 & -9.475 & 4U8T & -9.130 & -7.535 \\
	4Z31 & -8.790 & -9.061 & 5DNO & -7.830 & -8.431 & 5ELR & -6.250 & -8.894 \\
	5V7C & -6.990 & -8.994 & 5W1I & -12.790 & -10.833 & 5WZH & -10.460 & -10.535 \\
	6D12 & -9.300 & -10.184 & & &  & & &  \\
	
\bottomrule
\end{longtable}
}
	
\newpage	
\subsection{The S322 Dataset}

\renewcommand*{\arraystretch}{1.2}
\setlength{\tabcolsep}{6pt}

\centering
\begin{longtable}{%
		>{\centering\arraybackslash}p{1.5cm}
		>{\centering\arraybackslash}p{1.5cm}
		>{\centering\arraybackslash}p{1.5cm}
		>{\centering\arraybackslash}p{1.5cm}
		>{\centering\arraybackslash}p{1.5cm}
		>{\centering\arraybackslash}p{1.5cm}
		>{\centering\arraybackslash}p{1.5cm}
		>{\centering\arraybackslash}p{1.5cm}
		>{\centering\arraybackslash}p{1.5cm}}
	
\caption[Binding free energies for the PN dataset]%
{Experimental and predicted binding free energies ($\Delta G$) for the S322 dataset.\label{tab:S322}}\\
	
	\toprule
	\textbf{PDBID} & \textbf{Exp BA} & \textbf{Pred BA} &
	\textbf{PDBID} & \textbf{Exp BA} & \textbf{Pred BA} &
	\textbf{PDBID} & \textbf{Exp BA} & \textbf{Pred BA}\\
	\midrule
	\endfirsthead
	
	\toprule
	\textbf{PDBID} & \textbf{Exp BA} & \textbf{Pred BA} &
	\textbf{PDBID} & \textbf{Exp BA} & \textbf{Pred BA} &
	\textbf{PDBID} & \textbf{Exp BA} & \textbf{Pred BA}\\
	\midrule
	\endhead

	1HVO & -6.760 & -7.223 & 1WET & -7.940 & -11.499 & 2PUF & -7.769 & -10.978 \\
	1QPZ & -11.704 & -10.395 & 1QP0 & -10.165 & -10.829 & 1QP7 & -8.327 & -8.595 \\
	1FJX & -12.070 & -12.656 & 1J75 & -10.165 & -9.183 & 1JFS & -11.803 & -9.749 \\
	1J5K & -7.529 & -7.948 & 1PO6 & -9.276 & -9.960 & 1P51 & -11.528 & -11.533 \\
	1QZG & -8.614 & -9.556 & 1OSB & -9.754 & -10.591 & 1HVN & -7.227 & -6.756 \\
	1BDH & -8.242 & -8.219 & 1QP4 & -10.906 & -11.018 & 1QQB & -7.769 & -8.286 \\
	1TW8 & -12.192 & -12.227 & 1QQA & -8.722 & -8.317 & 1DH3 & -12.070 & -10.206 \\
	1JJ4 & -11.991 & -10.613 & 1FYM & -10.086 & -9.461 & 1JH9 & -11.859 & -7.513 \\
	1JT0 & -9.965 & -10.345 & 1P78 & -11.528 & -11.583 & 1P71 & -11.528 & -11.213 \\
	1OMH & -9.754 & -10.591 & 1S40 & -12.983 & -9.850 & 1U1L & -9.503 & -8.936 \\
	1U1R & -9.053 & -9.062 & 1U1N & -8.862 & -9.658 & 1U1O & -9.855 & -9.707 \\
	2BJC & -14.996 & -9.751 & 1ZZI & -7.713 & -8.013 & 2B0D & -8.242 & -9.262 \\
	2GII & -12.783 & -12.776 & 2GIE & -12.592 & -12.521 & 2GIH & -12.252 & -12.257 \\
	2ADY & -10.707 & -10.172 & 2AC0 & -10.364 & -10.105 & 2AYB & -11.859 & -11.788 \\
	2ERE & -10.256 & -10.373 & 2GE5 & -8.751 & -8.892 & 2O6M & -10.751 & -9.809 \\
	3D6Z & -8.204 & -6.798 & 2VYE & -9.991 & -9.824 & 3IGM & -8.590 & -8.917 \\
	3HXQ & -12.572 & -12.574 & 3EQT & -9.514 & -9.441 & 3H15 & -7.455 & -8.787 \\
	2KAE & -11.039 & -10.011 & 3MQ6 & -12.572 & -10.746 & 3JSP & -11.995 & -11.719 \\
	3AAF & -8.994 & -9.069 & 3N1L & -11.407 & -11.236 & 3N1K & -11.236 & -11.405 \\
	3R8F & -9.487 & -10.296 & 2RRA & -7.940 & -8.951 & 3QMI & -6.709 & -6.751 \\
	2KXN & -7.700 & -8.384 & 1U1P & -9.107 & -9.258 & 1U1M & -8.901 & -9.141 \\
	1U1K & -8.856 & -9.163 & 1U1Q & -9.706 & -9.535 & 2AOR & -8.242 & -8.350 \\
	2AOQ & -7.438 & -8.306 & 2I9K & -12.983 & -11.709 & 2GIJ & -12.783 & -12.774 \\
	2GIG & -12.252 & -12.259 & 2CCZ & -9.543 & -9.584 & 2AHI & -10.496 & -10.172 \\
	2ERG & -10.364 & -10.075 & 2AYG & -11.859 & -11.878 & 2ES2 & -8.843 & -8.857 \\
	2ATA & -9.646 & -10.318 & 2NP2 & -9.897 & -10.358 & 3D6Y & -6.799 & -8.204 \\
	2VY1 & -9.573 & -9.601 & 3HXO & -12.572 & -12.576 & 3GIB & -12.070 & -9.609 \\
	3D2W & -9.355 & -8.989 & 2KKF & -5.939 & -8.127 & 3M9E & -9.388 & -9.825 \\
	3JSO & -12.402 & -11.298 & 3K3R & -11.966 & -12.190 & 3N1I & -11.554 & -11.383 \\
	3N1J & -11.386 & -11.553 & 3KJP & -11.180 & -9.073 & 3Q0B & -8.134 & -9.216 \\
	3PIH & -11.890 & -10.759 & 3QMH & -6.502 & -6.837 & 3QMB & -7.438 & -6.471 \\
	3U7F & -8.482 & -8.392 & 3RN2 & -9.476 & -8.669 & 4ATK & -11.956 & -11.746 \\
	2LTT & -10.628 & -10.365 & 4HIO & -10.467 & -10.124 & 4HIK & -10.388 & -10.374 \\
	4A76 & -9.855 & -9.249 & 4HJ7 & -11.209 & -10.307 & 4HID & -8.273 & -10.406 \\
	4LJ0 & -8.590 & -9.679 & 4HT8 & -9.471 & -10.618 & 4GCK & -9.256 & -9.238 \\
	4F2J & -10.364 & -10.413 & 4GCT & -9.915 & -9.734 & 4NI7 & -13.223 & -10.421 \\
	4CH1 & -9.003 & -9.759 & 4QJU & -8.768 & -10.730 & 4ZBN & -13.194 & -10.140 \\
	4R55 & -7.876 & -9.748 & 4S0N & -10.751 & -9.187 & 4ZSF & -11.639 & -10.971 \\
	4RKG & -6.258 & -8.765 & 5A72 & -9.916 & -10.035 & 5DWA & -12.332 & -10.702 \\
	5K83 & -7.913 & -7.139 & 5ITH & -10.256 & -8.939 & 5W9S & -9.543 & -7.754 \\
	5YI3 & -8.534 & -9.757 & 5XFP & -8.072 & -8.756 & 5MEY & -9.265 & -9.248 \\
	5MEZ & -9.001 & -9.730 & 5W2M & -7.084 & -7.220 & 5K17 & -6.891 & -9.173 \\
	6MG1 & -10.276 & -11.093 & 5VMV & -12.769 & -9.678 & 6CNQ & -8.242 & -8.450 \\
	6FQP & -9.229 & -9.255 & 5WWF & -9.029 & -9.019 & 6FQQ & -8.739 & -9.723 \\
	5ZVB & -5.990 & -5.836 & 5ZVA & -5.809 & -6.133 & 5MPF & -9.605 & -9.268 \\
	6G1L & -12.332 & -11.436 & 6IIQ & -8.024 & -8.421 & 6IIR & -7.570 & -8.161 \\
	3ON0 & -11.341 & -10.307 & 4HQU & -14.586 & -11.704 & 4HQX & -12.162 & -13.473 \\
	4A75 & -9.232 & -9.772 & 3QSU & -10.511 & -9.455 & 4HJ8 & -10.440 & -8.520 \\
	4HIM & -10.132 & -10.460 & 4HJ5 & -9.817 & -10.383 & 4HP1 & -7.028 & -8.579 \\
	4J1J & -8.584 & -10.562 & 4NM6 & -8.140 & -8.975 & 4GCL & -9.278 & -9.652 \\
	3ZPL & -11.751 & -10.172 & 3ZH2 & -10.057 & -11.072 & 4HT4 & -11.039 & -10.384 \\
	4LNQ & -8.011 & -9.377 & 4LJR & -10.244 & -9.756 & 4R56 & -9.370 & -6.727 \\
	4TMU & -12.030 & -8.069 & 4R22 & -10.819 & -10.377 & 4Z3C & -7.661 & -8.490 \\
	3WPC & -10.496 & -11.373 & 3WPD & -11.619 & -10.158 & 2N8A & -9.573 & -10.047 \\
	5HRT & -12.114 & -9.964 & 5T1J & -10.526 & -10.033 & 5VC9 & -9.265 & -7.799 \\
	5HLG & -8.873 & -10.626 & 5W9Q & -8.482 & -7.439 & 6ASB & -7.940 & -9.214 \\
	6ASD & -7.661 & -8.398 & 5K07 & -6.122 & -8.960 & 5YI2 & -9.738 & -8.638 \\
	6MG3 & -11.449 & -10.346 & 6FWR & -8.873 & -9.202 & 6CNP & -8.391 & -8.260 \\
	5ZD4 & -10.798 & -9.287 & 5ZMO & -9.163 & -9.864 & 6BWY & -6.251 & -9.403 \\
	6CRM & -7.264 & -12.078 & 6CC8 & -7.181 & -9.082 & 6BUX & -5.807 & -7.718 \\
	6A2I & -8.701 & -7.413 & 5ZKI & -8.015 & -9.534 & 6KBS & -7.713 & -9.784 \\
	5ZKL & -10.479 & -10.217 & 5ZMD & -7.407 & -9.116 & 6ON0 & -9.605 & -9.809 \\
	2A9X & -9.303 & -10.954 & 3BX3 & -9.874 & -10.191 & 3IRW & -14.996 & -10.855 \\
	3IWN & -12.270 & -13.425 & 3K0J & -10.992 & -13.428 & 3QGB & -11.528 & -11.217 \\
	2XS2 & -10.113 & -7.906 & 3QGC & -10.982 & -11.066 & 4GHA & -8.590 & -9.489 \\
	3V6Y & -11.464 & -11.555 & 2LUP & -6.090 & -10.426 & 4M59 & -9.531 & -12.065 \\
	2MTV & -9.526 & -7.928 & 4RCM & -6.350 & -9.135 & 5EIM & -8.123 & -8.359 \\
	5GXH & -7.661 & -9.283 & 5KLA & -10.334 & -11.348 & 5EN1 & -8.942 & -8.923 \\
	5F5H & -9.668 & -9.273 & 5WZK & -9.265 & -10.222 & 5HO4 & -9.462 & -9.199 \\
	5U9B & -9.388 & -8.892 & 5WZJ & -10.751 & -10.287 & 5TF6 & -11.639 & -10.691 \\
	5SZE & -10.459 & -9.497 & 5WZG & -10.218 & -10.339 & 5M8I & -5.736 & -8.311 \\
	5UDZ & -9.592 & -9.344 & 5YTV & -8.043 & -7.952 & 6FQ3 & -8.015 & -8.838 \\
	5YTX & -8.007 & -7.982 & 5YKI & -12.030 & -9.948 & 6DCL & -10.647 & -8.916 \\
	6GD3 & -5.191 & -6.825 & 6G2K & -6.427 & -6.511 & 6CMN & -11.727 & -9.292 \\
	5YTS & -7.577 & -7.671 & 5WWG & -9.077 & -8.855 & 5YTT & -7.394 & -7.775 \\
	6FQR & -6.073 & -7.321 & 6GX6 & -7.264 & -6.222 & 5WWE & -8.550 & -8.906 \\
	6GC5 & -8.435 & -7.224 & 6RT6 & -6.755 & -7.762 & 6NOF & -10.617 & -10.431 \\
	6R7B & -10.558 & -9.540 & 6NOC & -10.490 & -10.469 & 6NOH & -10.421 & -10.387 \\
	6A6J & -9.068 & -8.487 & 6NOD & -10.280 & -10.526 & 6G99 & -6.176 & -9.008 \\
	6RT7 & -7.405 & -7.314 & 6U9X & -10.086 & -9.702 & 6NY5 & -9.952 & -10.192 \\
	6GBM & -5.556 & -9.573 & 1EC6 & -7.990 & -9.004 & 1M8Y & -9.270 & -12.497 \\
	1RPU & -13.300 & -9.771 & 1UTD & -16.890 & -9.030 & 2B6G & -10.600 & -10.012 \\
	2ERR & -12.200 & -9.110 & 2F8K & -10.300 & -9.927 & 2G4B & -7.830 & -8.468 \\
	2KFY & -8.840 & -8.126 & 2KG0 & -7.400 & -8.128 & 2KG1 & -7.450 & -8.131 \\
	2KX5 & -11.400 & -9.967 & 2L41 & -4.250 & -7.744 & 2LA5 & -11.500 & -10.760 \\
	2LEB & -9.320 & -8.734 & 2LEC & -9.440 & -8.680 & 2M8D & -8.730 & -8.680 \\
	2MJH & -9.670 & -8.771 & 2MXY & -7.900 & -6.914 & 2MZ1 & -6.870 & -7.871 \\
	2N82 & -10.300 & -10.363 & 2RU3 & -8.520 & -9.081 & 2XC7 & -7.040 & -8.859 \\
	2XFM & -8.240 & -7.836 & 2XNR & -5.400 & -7.462 & 2ZKO & -8.050 & -10.219 \\
	3BSB & -10.200 & -10.955 & 3BSX & -11.500 & -10.696 & 3BX2 & -9.970 & -9.730 \\
	3K49 & -12.050 & -10.952 & 3K4E & -10.800 & -11.859 & 3K5Q & -9.300 & -9.072 \\
	3K5Y & -10.540 & -9.511 & 3K5Z & -9.380 & -9.682 & 3K61 & -8.890 & -9.531 \\
	3K62 & -8.740 & -9.253 & 3K64 & -9.220 & -8.886 & 3L25 & -8.175 & -9.260 \\
	3LQX & -12.220 & -10.909 & 3MDG & -7.850 & -8.917 & 3MOJ & -13.000 & -10.957 \\
	3NCU & -10.300 & -9.814 & 3NMR & -8.440 & -7.299 & 3NNH & -6.190 & -8.698 \\
	3O3I & -6.520 & -7.070 & 3O6E & -7.070 & -6.524 & 3Q0L & -12.300 & -11.392 \\
	3Q0M & -12.500 & -12.818 & 3Q0N & -10.500 & -10.804 & 3Q0P & -12.800 & -10.043 \\
	3Q0Q & -13.300 & -12.542 & 3Q0R & -13.800 & -12.954 & 3Q0S & -10.800 & -11.713 \\
	3QG9 & -10.600 & -11.263 & 3U4M & -15.900 & -10.085 & 3V71 & -11.100 & -10.371 \\
	3V74 & -11.600 & -11.382 & 3WBM & -9.980 & -10.822 & 4CIO & -9.760 & -9.002 \\
	4ED5 & -9.080 & -8.516 & 4ERD & -9.540 & -10.329 & 4JVH & -9.590 & -9.484 \\
	4KJI & -8.940 & -10.146 & 4LG2 & -9.540 & -8.843 & 4NL3 & -7.380 & -10.077 \\
	4O26 & -8.200 & -11.145 & 4OE1 & -12.110 & -9.541 & 4QI2 & -9.440 & -9.828 \\
	4QVC & -11.560 & -9.994 & 4QVD & -10.120 & -11.377 & 4R3I & -7.770 & -6.765 \\
	4RCJ & -8.180 & -8.440 & 4TUX & -10.400 & -9.287 & 4U8T & -9.130 & -7.841 \\
	4Z31 & -8.790 & -9.306 & 5DNO & -7.830 & -8.538 & 5ELR & -6.250 & -8.793 \\
	5V7C & -6.990 & -9.070 & 5W1I & -12.790 & -10.243 & 5WZH & -10.460 & -10.308 \\
	6D12 & -9.300 & -10.271 & & &  & & &  \\
	
	\bottomrule
\end{longtable}
\newpage		
	
	\bibliography{SI}